\documentclass[preprintnumbers]{revtex4}

\usepackage{graphicx}
\def\Re{{\cal R \mskip-4mu \lower.1ex \hbox{\it e}\,}}
\def\Im{{\cal I \mskip-5mu \lower.1ex \hbox{\it m}\,}}
\def\ie{{\it i.e.}}
\def\eg{{\it e.g.}}

\def\etal{{\it et al.}}

\def\mpl{\ifmmode \overline M_{Pl}\else $\overline M_{Pl}$\fi}
\def\sub#1{_{\lower.25ex\hbox{$\scriptstyle#1$}}}
\def\tev{\,{\ifmmode\mathrm {TeV}\else TeV\fi}}
\def\gev{\,{\ifmmode\mathrm {GeV}\else GeV\fi}}
\def\mev{\,{\ifmmode\mathrm {MeV}\else MeV\fi}}
\def\to{\rightarrow}

\def\subw{_{\rm w}}
\def\mh{\ifmmode m\sbl H \else $m\sbl H$\fi}
\def\mch{\ifmmode m_{H^\pm} \else $m_{H^\pm}$\fi}
\def\mt{\ifmmode m_t\else $m_t$\fi}
\def\mc{\ifmmode m_c\else $m_c$\fi}
\def\mz{\ifmmode M_Z\else $M_Z$\fi}
\def\mw{\ifmmode M_W\else $M_W$\fi}
\def\mws{\ifmmode M_W^2 \else $M_W^2$\fi}
\def\mhs{\ifmmode m_H^2 \else $m_H^2$\fi}   
\def\mzs{\ifmmode M_Z^2 \else $M_Z^2$\fi}
\def\mts{\ifmmode m_t^2 \else $m_t^2$\fi}
\def\mcs{\ifmmode m_c^2 \else $m_c^2$\fi}
\def\mchs{\ifmmode m_{H^\pm}^2 \else $m_{H^\pm}^2$\fi}
\def\ztwo{\ifmmode Z_2\else $Z_2$\fi}
\def\zone{\ifmmode Z_1\else $Z_1$\fi}
\def\mtwo{\ifmmode M_2\else $M_2$\fi}
\def\mone{\ifmmode M_1\else $M_1$\fi}
\def\tb{\ifmmode \tan\beta \else $\tan\beta$\fi}
\def\xw{\ifmmode x\subw\else $x\subw$\fi}
\def\ch{\ifmmode H^\pm \else $H^\pm$\fi}
\def\lum{\ifmmode {\cal L}\else ${\cal L}$\fi}
\def\inpb{\,{\ifmmode {\mathrm {pb}}^{-1}\else ${\mathrm {pb}}^{-1}$\fi}}
\def\infb{\,{\ifmmode {\mathrm {fb}}^{-1}\else ${\mathrm {fb}}^{-1}$\fi}}
\def\epem{\ifmmode e^+e^-\else $e^+e^-$\fi}
\def\ppb{\ifmmode \bar pp\else $\bar pp$\fi}
\def\bsg{\ifmmode B\to X_s\gamma\else $B\to X_s\gamma$\fi}
\def\bsll{\ifmmode B\to X_s\ell^+\ell^-\else $B\to X_s\ell^+\ell^-$\fi}
\def\bstt{\ifmmode B\to X_s\tau^+\tau^-\else $B\to X_s\tau^+\tau^-$\fi}
\def\lamt{\ifmmode \tilde\lambda\else $\tilde\lambda$\fi}
\def\shat{\ifmmode \hat s\else $\hat s$\fi}
\def\that{\ifmmode \hat t\else $\hat t$\fi}
\def\uhat{\ifmmode \hat u\else $\hat u$\fi}

\def\matth{\mathsurround=0pt}

\def\atversim#1#2{\lower0.7ex\vbox{\baselineskip\zatskip\lineskip\zatskip
  \lineskiplimit 0pt\ialign{$\matth#1\hfil##\hfil$\crcr#2\crcr\sim\crcr}}}

\def\ie{{\it i.e.}}
\def\eg{{\it e.g.}}

\def\etal{{\it et al.}}

\def\mpl{\ifmmode \overline M_{Pl}\else $\overline M_{Pl}$\fi}
\def\to{\rightarrow}

\begin{document}
\bibliographystyle{revtex}

\preprint{SLAC-PUB-9184}

\title{Signals for Noncommutative QED at Future $e^+e^-$ Colliders}

\author{Thomas G. Rizzo}

\email[]{rizzo@slac.stanford.edu}
\affiliation{Stanford Linear Accelerator Center, 
Stanford University, Stanford, California 94309 USA}

\date{\today}

\begin{abstract}
The signatures for noncommutative QED at $e^+e^-$ colliders with center of 
mass energies in the range of 0.5-5 TeV are examined. For integrated 
luminosities of 0.5-1 ab$^{-1}$ or more, sensitivities to the associated mass 
scales greater than $\sqrt s$ are possible. 
\end{abstract}

\maketitle

\section{Introduction}

If the Planck scale is indeed of order a few TeV, some potential stringy
effects may be observable at future colliders in addition to the existence of
extra dimensions. One such possibility is that near the string scale
space-time becomes noncommutative (NC), \ie, the co-ordinates themselves no
longer commute: $[x_\mu,x_\nu]=i\theta_{\mu\nu}$,
where the $\theta_{\mu\nu}$ are a constant, frame-independent set of six
dimensionful parameters{\cite {rev}}.  This may occur in string theory in
the presence of background fields.  The
$\theta_{\mu\nu}$ may be separated into two classes: (i) space-space
noncommutivity with  $\theta_{ij}=c_{ij}/\Lambda_B^2$, and (ii) space-time
noncommutivity with $\theta_{0i}=c_{0i}/\Lambda_E^2$.
The auxiliary quantities 
$(\hat c_E)_i=c_{0i}$ and $(\hat c_B)_k=\epsilon_{ijk}c_{ij}$, with
$ijk$ cyclic, can also be defined, where $\hat c_{E,B}$ are two, fixed,
frame-independent unit
vectors associated with the mass scales $\Lambda_{E,B}$.  These NC scales are
anticipated to lie above a TeV and the influence of NC physics 
will only become apparent as the these scales are approached.
Since the commutator of the co-ordinates is 
{\it not} a tensor and is frame-independent, NC theories violate Lorentz 
invariance (but can be shown to conserve CPT), 
with the two vectors $\hat c_{E,B}$ 
being preferred directions in space, related to the directions of the
background fields.   We note that 
momenta still commute in the usual way, thus energy and momentum remain 
conserved quantities.  Since experimental 
probes of NC theories are sensitive 
to the directions of $\hat c_{E,B}$, experiments must 
employ astronomical coordinate systems and time-stamp their data so that, 
\eg, the rotation of the Earth or the Earth's motion around the Sun does not 
wash out or dilute the effect through time-averaging.

\begin{figure}[htbp]
\centerline{
\includegraphics[width=6.3cm,angle=90]{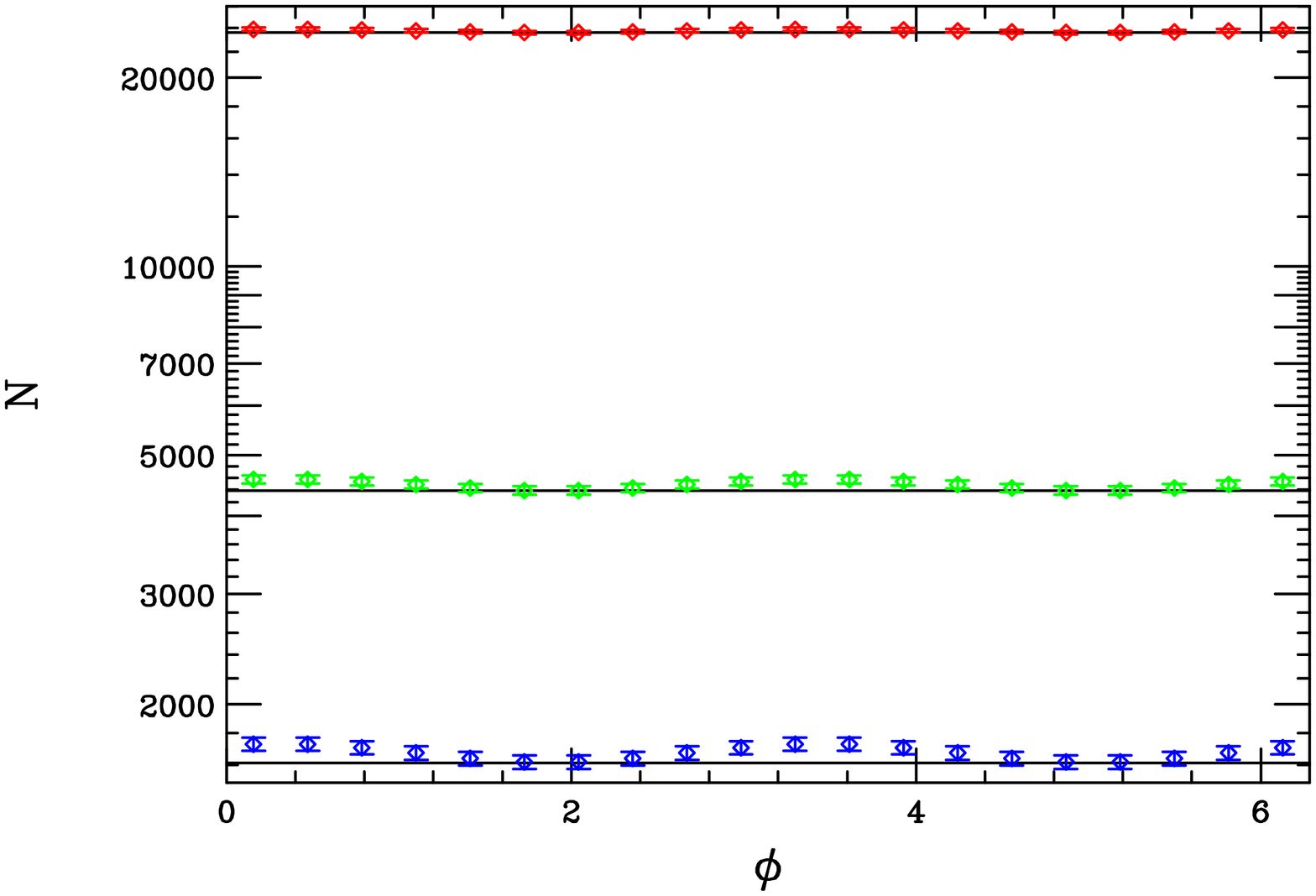}}
\vspace*{5mm}
\centerline{
\includegraphics[width=6.3cm,angle=90]{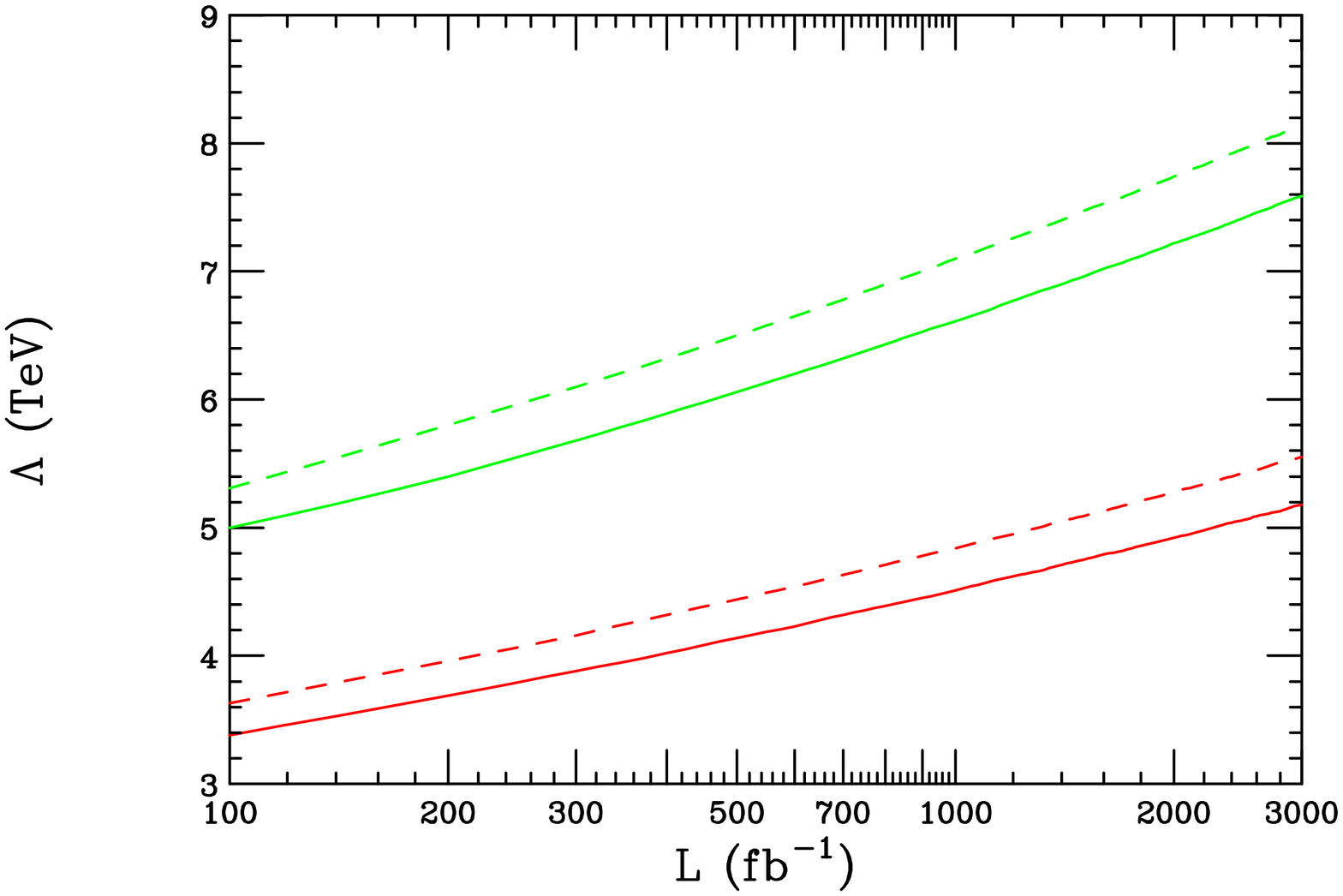}}
\vspace*{0.1cm}
\caption{(Top) Binned $\phi$ distribution for Bhabha scattering at a 3 TeV 
CLIC assuming an integrated luminosity of $1~ab^{-1}$ with 
$|\cos \theta|$ cuts of 0.9, 0.7 and 0.5 (from top to bottom). The solid line 
is the SM prediction while the data assumes $c_{02}=1$ and $\Lambda_E=3$ TeV. 
(Bottom) Sensitivity to the scale $\Lambda_E$ at a $\sqrt s=3(5)$ TeV 
CLIC corresponding to the lower (upper) set of curves. The dashed (solid) 
curve is for the case $c_{01} (c_{02})=1$. }
\label{E3064_fig1}
\end{figure}

It is possible to construct noncommutative analogs of conventional field 
theories following either the 
Weyl-Moyal (WM){\cite {moyal}} or 
Seiberg-Witten (SW){\cite {sw}} approaches, both of which have their own 
advantages and disadvantages. In the SW approach, the field theory 
is expanded in a power series in $\theta$  which 
then produces an infinite tower of additional operators. At any fixed order 
in $\theta$, the theory can be shown{\cite {nonr}} to be non-renormalizable. 
The SW construction can, however, be applied to any gauge theory 
with arbitrary matter representations. 
In the WM approach, only $U(N)$ gauge theories are found to be closed under 
the group algebra and the matter content is restricted to the 
(anti-)fundamental and adjoint representations. Further restrictions on matter 
representations apply when a product of group factors is present, such as 
in the SM{\cite {jab}}. These theories are at least  
one-loop renormalizable and appear to remain so even when 
spontaneously broken{\cite {renorm}}.

\begin{figure}[htbp]
\centerline{
\includegraphics[width=6.3cm,angle=90]{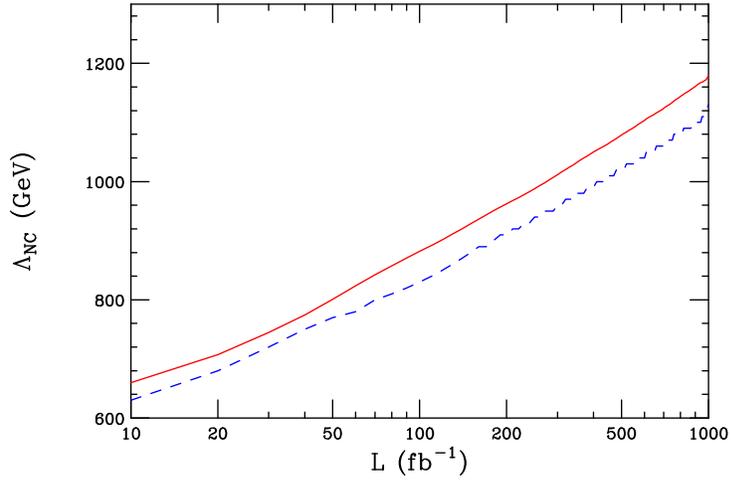}}
\vspace*{0.1cm}
\caption{Same as the bottom portion of  Fig. 1 except now for a 
$\sqrt s$=500 GeV collider and the labels on the curves reversed.}
\label{E3064_fig1p}
\end{figure}

These distinctive properties of NC gauge theories render the construction 
of a satisfactory noncommutative version of the Standard Model (SM), the NCSM, 
quite difficult{\cite {models}}. However, NCQED is a well-defined theory in 
the WM approach and we explore here its implications for very high energy 
$e^+e^-$ colliders following this prescription. 
This version of NCQED differs from ordinary QED in 
several ways: ($i$) the $ee\gamma$ vertex picks up a Lorentz violating 
phase factor given by $e^{i\theta_{\mu\nu}p_i^\mu p_o^\nu /2}$ 
where $p_i$ and $p_o$ are the incoming and outgoing electron momenta; 
($ii$) the  
NC theory predicts trilinear and quartic couplings between the photons that 
are, to leading order, linear and quadratic in the parameters 
$\theta_{\mu\nu}$, respectively, and are kinematics dependent; ($iii$) only 
the charges 
$Q=0,\pm 1$ are allowed by gauge invariance in NCQED{\cite {haya}}. 
Thus quarks cannot be accommodated in the theory as it presently exists 
and an extension to a full 
NCSM is required. We note that these difficulties can apparently be 
circumvented using the SW approach if one is willing to pay the price of 
nonrenormalizanlity and ambiquities in multiple gauge boson couplings. 
This restriction to $Q=0,\pm 1$ implies that we can only examine the NC 
effects for 
the handful of processes which have external charged leptons or photons. 
Despite this limitation,
NCQED provides a testing ground for the basic ideas 
behind NC quantum field theory
and has had its phenomenological implications examined by a 
number of authors{\cite {ncqed}}. As we will see, the hallmark signal at 
colliders for NCQED is the appearance of an azimuthally-dependent cross 
section in $2\to 2$ processes; the azimuthal dependence arises from the 
existence of the two preferred directions discussed above. We note that 
in NC gauge theories, 
the leading NC contributions can be shown to take the form of 
new dimension-8 operators whose scale is set by $\Lambda_{E,B}$. 

\begin{figure}[htbp]
\centerline{
\includegraphics[width=6.3cm,angle=90]{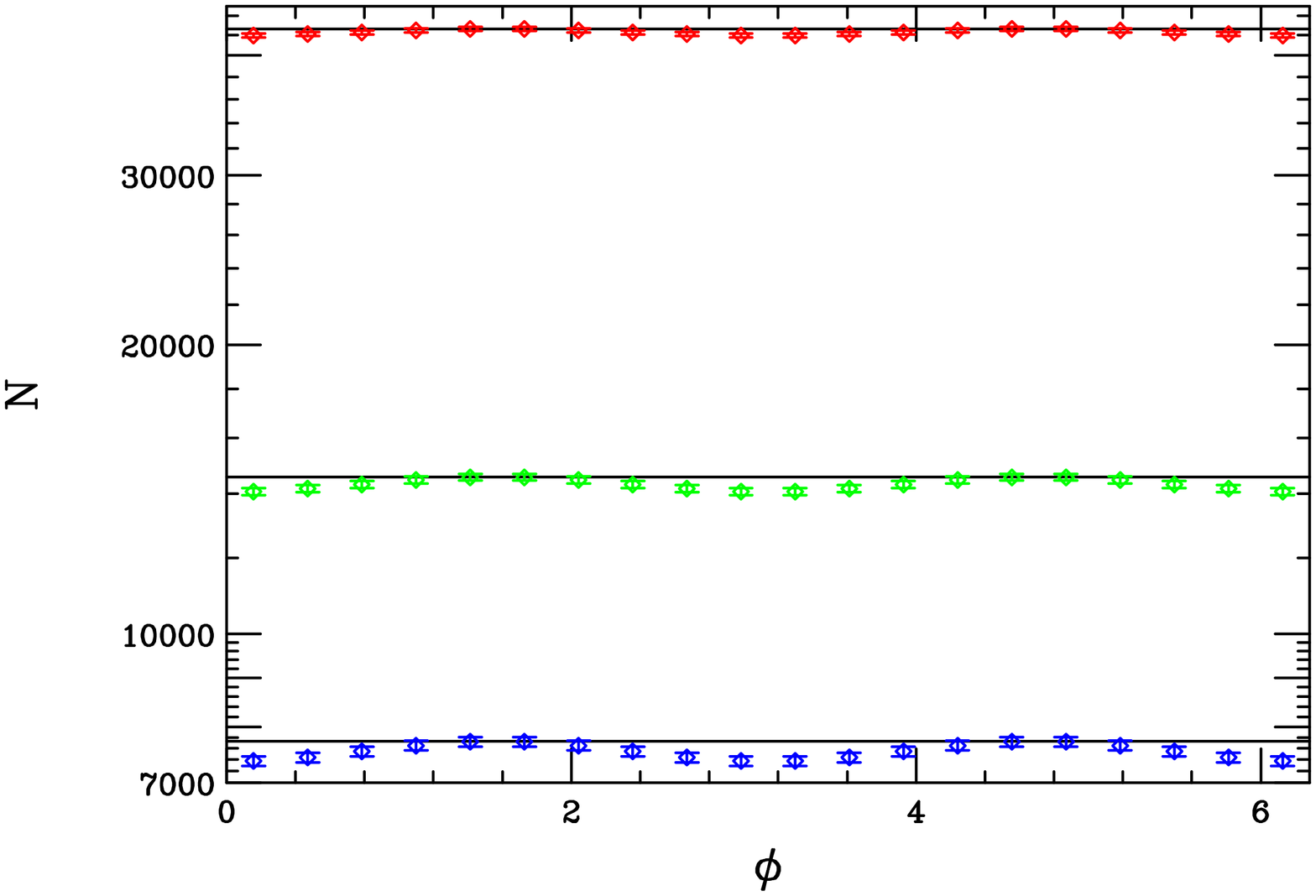}}
\vspace*{5mm}
\centerline{
\includegraphics[width=6.3cm,angle=90]{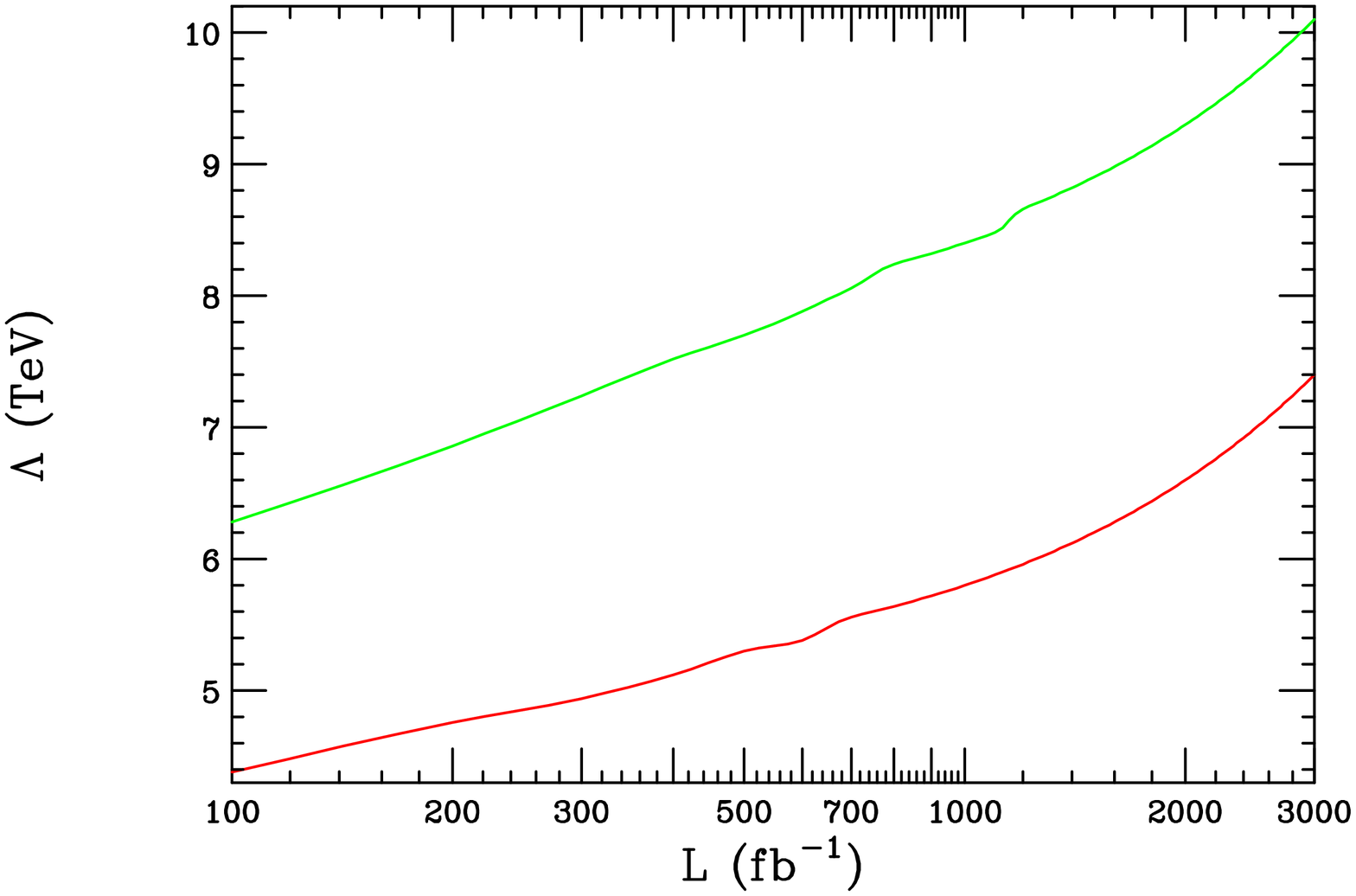}}
\vspace*{0.1cm}
\caption{(Top) Same as in Fig.1  but now for M\"oller scattering 
assuming $c_{12}=1$. 
(Bottom) Sensitivity to the scale $\Lambda_B$ at a $\sqrt s=3 (5)$ TeV 
CLIC corresponding to the lower (upper) curves.}
\label{E3064_fig1b}
\end{figure}

A caveat to our analysis below is the question of how/if 
the SM couplings of the $Z$ boson to $e^+e^-$ are modified in the NC case, as
they contribute to  
Bhabha and M\"oller scattering. In truth, this lies outside the realm of the 
NCQED model and can only be addressed within a full NCSM. Here we will assume 
that these $Z$ couplings get rescaled by kinematic-dependent exponents in a 
manner identical to photons. Within the SW approach we know that this is 
indeed what happens for on-shell electrons, to leading order in 
$\theta_{\mu\nu}$, and we might expect it to remain true in higher orders in 
the complete NCSM theory.

\begin{figure}[htbp]
\centerline{
\includegraphics[width=6.3cm,angle=90]{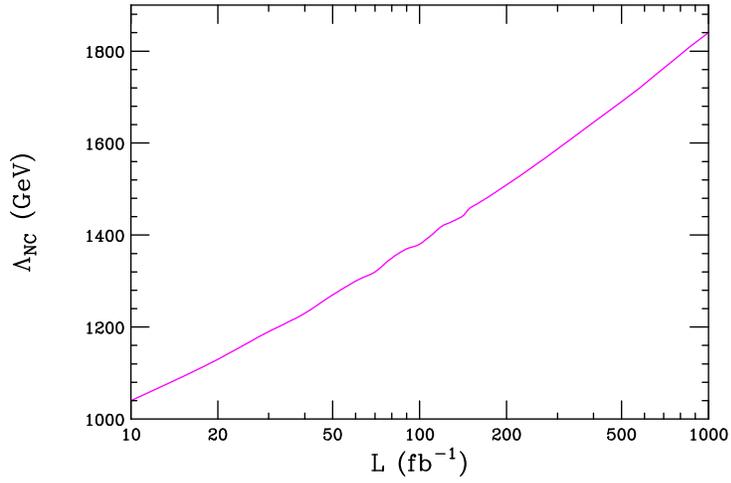}}
\vspace*{0.1cm}
\caption{Same as in the bottom portion of Fig.2 but now for 
M\"oller scattering assuming $\sqrt s$=500 GeV and $c_{12}=1$.}
\label{E3064_fig1bp}
\end{figure}
\begin{figure}[htbp]
\centerline{
\includegraphics[width=8.7cm,angle=0]{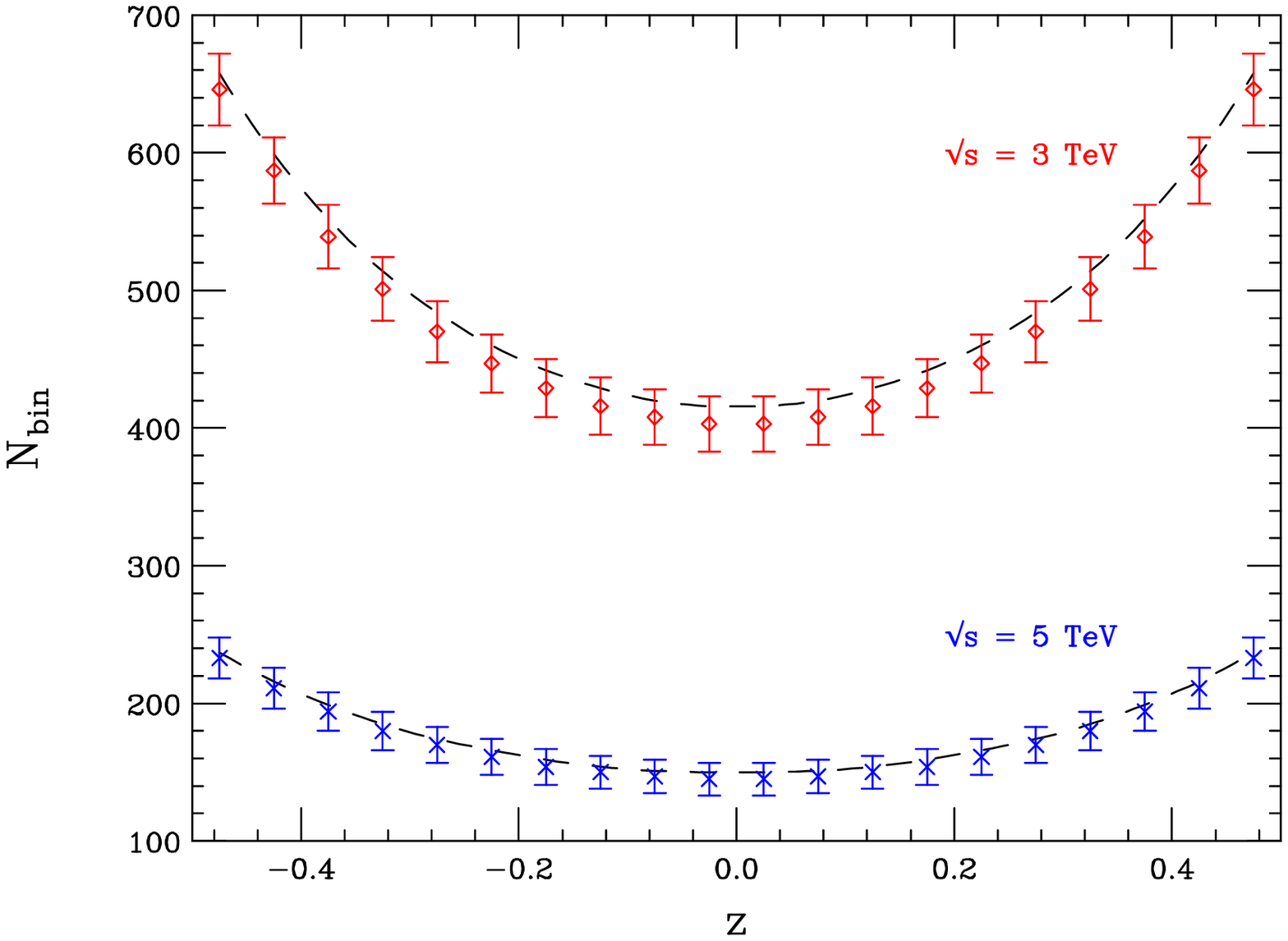}}
\vspace*{5mm}
\centerline{
\includegraphics[width=8.7cm,angle=0]{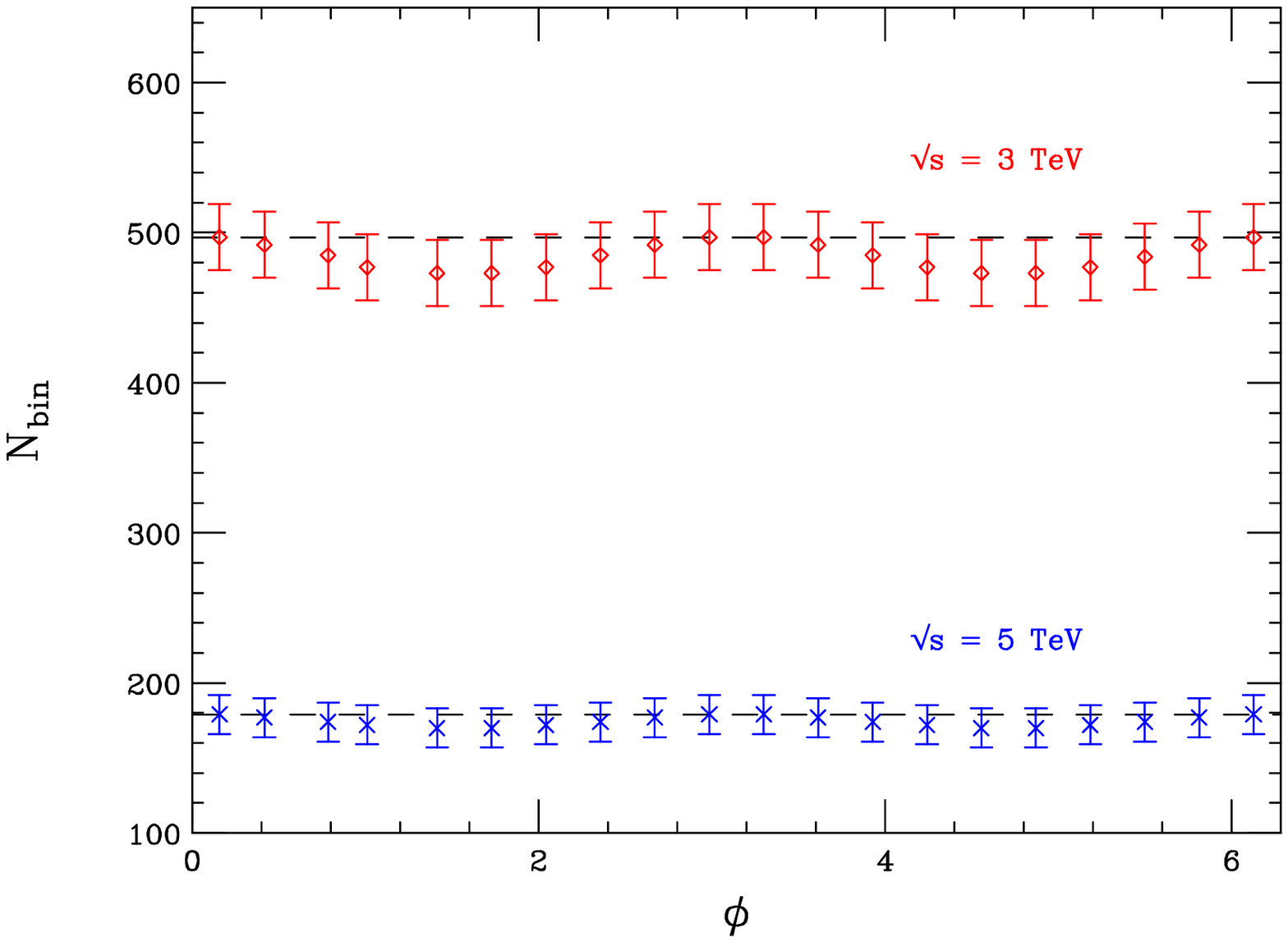}}
\vspace*{0.1cm}
\caption{Shifts in the $z=\cos \theta$ and $\phi$ distributions for the 
process $e^+e^- \to \gamma \gamma$ at a 3 or 5 
TeV CLIC assuming an integrated luminosity of 1 ab$^{-1}$. The dashed curves 
show the SM expectations while the `data' assumes $c_{02}=1$ and 
$\Lambda_E=\sqrt s$. A cut of $|z|<0.5$ has been applied in the $\phi$
distribution.}
\label{E3064_fig2}
\end{figure}

At typical linear collider energies, \ie, 500-100 GeV with integrated 
luminosities in the range approaching 500 fb$^{-1}$, NC physics can be 
probed at the TeV scale.
Very high energy $e^+e^-$ colliders such as CLIC will allow us to examine 
values of $\Lambda_{E,B}$ beyond  
several TeV provided sufficient luminosity, $\sim 1$ ab$^{-1}$,  
is available. To demonstrate these claims we will examine three specific 
processes: Bhabha and  M\"oller scattering, as well as pair annihilation. 
In all cases we assume the incoming $e^-$ direction 
to be along the $z-$axis.

\section{Results}

The first process we consider is Bhabha scattering which proceeds through 
both $s-$ and $t-$channel gauge boson exchanges. There are no new amplitudes
to consider in this case, 
but each vertex picks up the kinematic dependent phase discussed above. 
As demonstrated{\cite {ncqed}} in 
earlier work, the NC modifications to the SM result only appear in 
the interference term in the squared matrix element and are sensitive only to 
finite $\Lambda_E$. This NC effect appears through the cosine of the relative 
phase:  $\Delta_{Bhabha}=\phi_s-\phi_t={-1\over {\Lambda_E^2}}
[c_{01}t+{\sqrt {ut}}(c_{02}c_\phi+c_{03}s_\phi)]$, where 
$t$ and $u$ are the usual Mandelstam variables.
Independently of the particular values of $c_{0i}$, the resulting $\cos \theta$ 
distribution is modified; however, for $c_{02}$ and/or $c_{03}$ 
nonvanishing, the cross section also picks up a periodic 
azimuthal (\ie, $\phi$) dependence as is shown in Fig.~\ref {E3064_fig1}. 
While other new physics may lead to modifications of the $\cos \theta$ 
distribution, only Lorentz violation
can induce a $\phi-$dependence in the 
cross section. Note that the $\phi$ 
dependence becomes more pronounced as stiffer cuts on the scattering 
angle, which selects more central events where both $u$ and $t$ are large,
are applied.  We find that  beam polarization is not particularly useful
for NC searches in Bhabha scattering; 
the Left-Right Polarization asymmetry, $A_{LR}$, 
is found to be rather insensitive to NC effects showing, in particular, 
little azimuthal 
dependence. The reach for $\Lambda_{NC}$ in this case for CLIC energies 
is also presented 
in Figure~\ref {E3064_fig1} where we see that it is of order $\sim 1.5\sqrt s$. 
Figure~2 shows that similar results are obtained at lower 
energy colliders. 

The next reaction we examine is M\"oller scattering. As in the case of
Bhabha scattering, 
the NC effects appear only in the cosine of the phase of the 
interference term between the two 
$t-$ and $u-$channel amplitudes : $\Delta_{Moller}=\phi_u-\phi_t=
{-{\sqrt {ut}}\over {\Lambda_B^2}}[c_{12}c_\phi-c_{31}s_\phi]$. Note that, 
unlike Bhabha scattering, M\"oller scattering is sensitive to 
a finite $\Lambda_B$. If either $c_{12}$ or $c_{31}$ is nonzero, an azimuthal 
dependence is seen in the cross section as 
displayed in Fig.~3. 
(Note that there is no sensitivity to a nonvanishing $c_{23}$.) The oscillatory 
behavior is somewhat more pronounced here than in Bhabha scattering. 
Again, no additional 
sensitivity arises from the azimuthal dependence of $A_{LR}$. The reach for 
$\Lambda_{NC}$ from M\"oller scattering is seen from Fig.~\ref {E3064_fig2} 
to exceed that for Bhabha scattering. 
Figure~4 shows that similar results are obtained at lower 
energy colliders. 

\begin{figure}[htbp]
\centerline{
\includegraphics[width=8.7cm,angle=0]{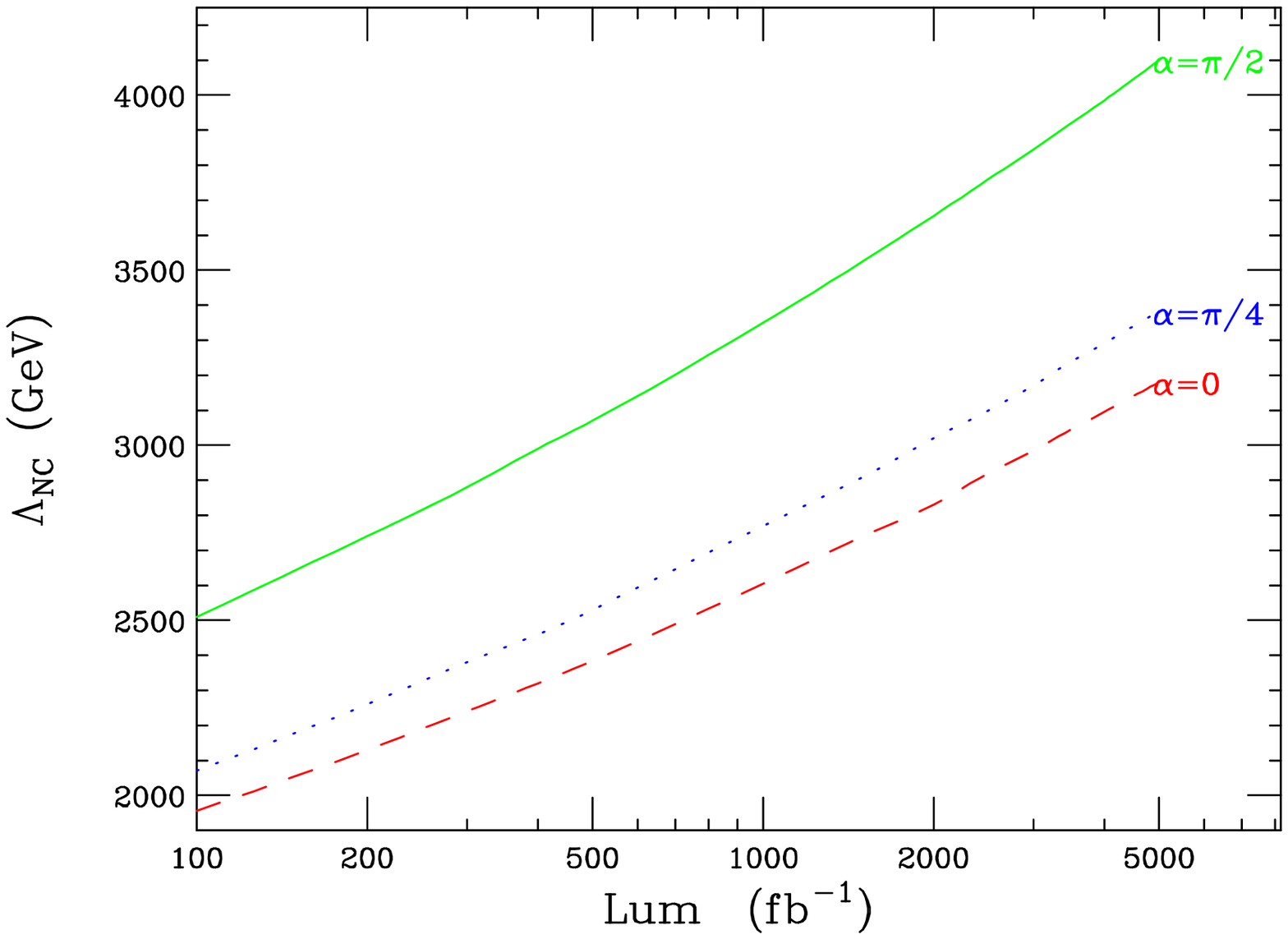}}
\vspace*{5mm}
\centerline{
\includegraphics[width=8.7cm,angle=0]{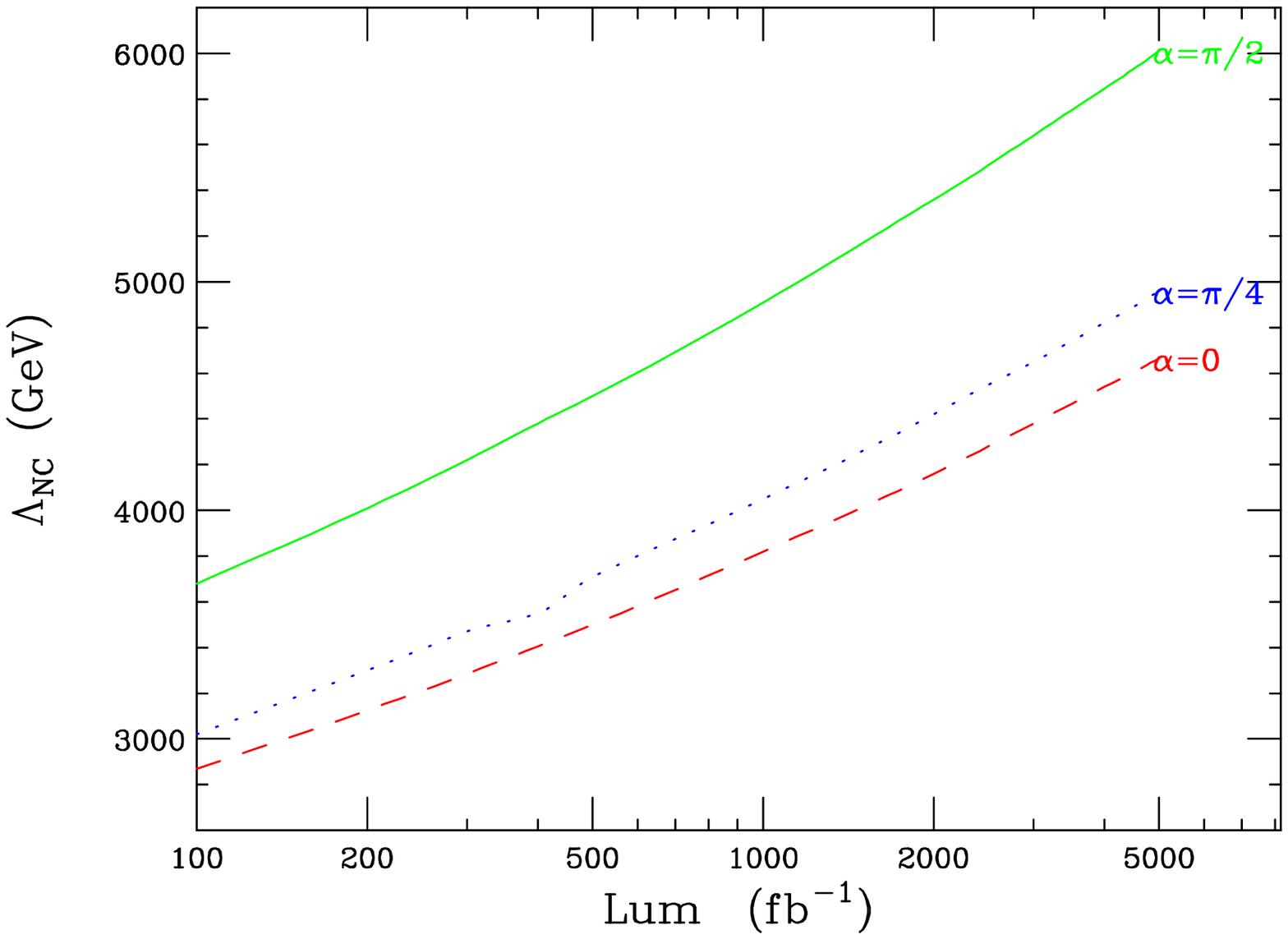}}
\vspace*{0.1cm}
\caption{Reach for $\Lambda_E$ at a (top) 3 TeV or a (bottom) 5 TeV CLIC as a 
function of the integrated luminosity for the process 
$e^+e^- \to \gamma \gamma$ following the notation in the text.}
\label{E3064_fig3}
\end{figure}

The last case we consider is pair annihilation. In addition to the new phases 
that enter the $t-$ and $u-$ channel amplitudes, there is now an
additional $s-$channel 
photon exchange graph involving the NC-generated three photon vertex 
discussed above. This new amplitude is proportional to the sine of the phase: 
$\Delta_{PA}= {{-s}\over {2 \Lambda^2_E}} [c_{01} c_\theta +c_{02} s_\theta 
c_\phi + c_{03} s_\theta s_\phi]$. 
As in Bhabha scattering, this modification  to the 
SM result appears to lowest order as a dimension-8 operator that probes finite 
$\Lambda_E$. As before,  the $\cos \theta$ 
distribution is modified for all values of the $c_{0i}$,
but a nonvanishing $c_{02}$ and/or $c_{03}$ is required to 
produce an azimuthal dependence. The resulting angular distributions are shown 
in Fig.~5 for the case $c_{02}=1$. Writing 
$c_{01}=\cos \alpha$ and $c_{02}=\sin \alpha \cos \beta$, and 
$c_{03}=\sin \alpha \sin \beta$, Fig.~6 displays
the reach for $\Lambda_E$ for CLIC energies for several values of $\alpha$.

In summary, we have examined the effects of NCQED in several $2\to 2$
scattering processes at high energy $e^+e^-$ colliders.  We find that
these effects produce an azimuthal dependence in the cross sections,
providing a unique signature of the Lorentz violation inherent in
these theories.  The search reaches for the NC scale in a variety
of processes are summarized in Table 1 for both $\sqrt s=500$ GeV 
and CLIC energies. We display the $95\%$ CL search limits on the NC scale
$\Lambda_{NC}$ from the various processes considered above at
a 500 GeV $e^+e^-$ linear collider with an integrated luminosity
of 500 fb$^{-1}$ or at a 3 or 5 TeV CLIC with an integrated luminosity 
of 1 ab$^{-1}$.  The sensitivities are from the first two papers 
in \cite{ncqed}. The $\gamma\gamma\to e^+e^-$ and $\gamma e\to\gamma e$ 
analyses of Godfrey and Doncheski include an overall $2\%$ systematic error 
not included by Hewett, Petriello and Rizzo.
We see that these machines have a reasonable sensitivity 
to NC effects and provide a good probe of such theories. 

\begin{table}
\caption{Summary of the $95\%$ CL search limits on the NC scale
$\Lambda_{NC}$.}
\centering
\begin{tabular}{|c|c|c|c|c|} \hline\hline
Process & Structure Probed & $\sqrt s$=500 GeV & $\sqrt s$=3 TeV & 
$\sqrt s$=5 TeV  \\ \hline
$\epem\to\gamma\gamma$   & Space-Time & $740-840$ GeV & $2.5-3.5$ TeV & 
$3.8-5.0$ TeV \\
Moller Scattering & Space-Space &  1700 GeV & 5.9 TeV & 8.5 TeV \\ 
Bhabha Scattering & Space-Time & 1050 GeV & $4.5-5.0$ TeV & $6.6-7.2$ TeV \\
$\gamma\gamma\to\gamma\gamma$ & Space-Time & $700-800$ GeV  & & \\ 
   & Space-Space & 500 GeV & & \\ 
$\gamma\gamma\to e^+e^-$ & Space-Time & $220-260$ GeV & $1.1-1.3$ TeV & 
$1.8-2.1$ TeV \\ 
$\gamma e\to\gamma e$ & Space-Time & $540-600$ GeV & $3.1-3.4$ TeV & 
$4.8-5.8$ TeV \\ 
 & Space-Space & $700-720$ GeV & $4.0-4.2$ TeV & $6.3-6.5$ TeV \\ \hline\hline
\end{tabular}
\label{summ}
\end{table}

%
\def\MPL #1 #2 #3 {Mod. Phys. Lett. {\bf#1},\ #2 (#3)}
\def\NPB #1 #2 #3 {Nucl. Phys. {\bf#1},\ #2 (#3)}
\def\PLB #1 #2 #3 {Phys. Lett. {\bf#1},\ #2 (#3)}
\def\PR #1 #2 #3 {Phys. Rep. {\bf#1},\ #2 (#3)}
\def\PRD #1 #2 #3 {Phys. Rev. {\bf#1},\ #2 (#3)}
\def\PRL #1 #2 #3 {Phys. Rev. Lett. {\bf#1},\ #2 (#3)}
\def\RMP #1 #2 #3 {Rev. Mod. Phys. {\bf#1},\ #2 (#3)}
\def\NIM #1 #2 #3 {Nuc. Inst. Meth. {\bf#1},\ #2 (#3)}
\def\ZPC #1 #2 #3 {Z. Phys. {\bf#1},\ #2 (#3)}
\def\EJPC #1 #2 #3 {E. Phys. J. {\bf#1},\ #2 (#3)}
\def\IJMP #1 #2 #3 {Int. J. Mod. Phys. {\bf#1},\ #2 (#3)}
\def\JHEP #1 #2 #3 {J. High En. Phys. {\bf#1},\ #2 (#3)}

\end{document}